\documentclass[final,5p,times,twocolumn,authoryear]{elsarticle}

\usepackage{amssymb}
\usepackage{amsmath}	
\usepackage{hyperref}
\usepackage{float} 
\usepackage{listings}

\journal{Journal of Quantitative Spectroscopy and Radiative Transfer}

\begin{document}

\begin{frontmatter}



\title{A numerically stable comoving frame solver for line radiative transfer}


\author[IvS]{T. Ceulemans}
\author[IvS]{F. De Ceuster}
\author[IvS]{L. Decin}

\affiliation[IvS]{organization={Institute of Astronomy},
	addressline={Celestijnenlaan 200D},
	city={Leuven},
	postcode={3001},
	country={Belgium}}	

\begin{abstract}
Radiative transfer is essential in astronomy, both for interpreting observations and simulating various astrophysical phenomena.
However, self-consistent line radiative transfer is computationally expensive, especially in 3D.
To reduce the computational cost when utilizing a discrete angular discretization, we use a comoving frame interpretation of the radiative transfer equation. The main innovation of this paper lies in the novel stabilization method for the resulting numerical discretization. The stabilization method is able to reduce spurious oscillatory behavior in the computed intensities, at the expense of extra boundary conditions which need to be enforced. We also implement an adaptive angular discretization for the ray-tracing implementation, in order to efficiently and accurately calculate the radiation field. Finally, we apply this new numerical method to compute NLTE line radiative transfer on a hydrodynamics model, showcasing its potential improvement in computation efficiency.
\end{abstract}



\begin{keyword}
Radiative transfer \sep Software: development \sep Numerical methods


\end{keyword}

\end{frontmatter}


\section{Introduction}\label{section: Introduction}

In the astronomy literature, various algorithms already exist for computing intensities in (line) radiative transfer, including discretizing the formal solution \citep[as in][]{auer_formal_2003}, using the Feautrier method \citep[][]{feautrier_sur_1964}, and moment method \citep[see e.g.][]{auer_use_1970}).
Self-consistent non-local thermodynamical equilibrium (NLTE) line radiative transfer comprises of a set of non-linear equations, in which the radiation field and the properties of the medium depend on each other. To solve this problem, one can either resort to an iterative approach, such as the accelerated lambda iteration of \cite{rybicki_accelerated_1991} or linearizing the equations \citep[as done in][]{auer_non-lte_1969}.
In this paper, we focus on the former approach. In this method, the slowest calculation step consists of computing the mean line intensity, which requires us to compute the radiation field at frequencies around the line center in the comoving frame at each position. To compute the intensity in a single direction, we require the computed intensity in the same direction from the upwind point.
Therefore, theoretically, we can optimize our computations by reusing previously computed intensity information, instead of restarting the intensity computation from the model boundary.
However, in case of a non-constant velocity field, we need to take into account the Doppler shift, which causes the required frequency ranges to be misaligned at the different positions.

To address this issue, one can rewrite the radiative transfer equation to the comoving frame following the approach outlined in \cite{hauschildt_fast_1992}, or \cite{castor_radiation_2004}, Chapter 6. However, these initial approaches were restricted to using monotonic velocity fields. In \cite{baron_co-moving_2004, hauschildt_improved_2004}, the authors proposed a new solution method, which also allows for a non-monotonic velocity field. Nonetheless, the solution method may encounter numerical instabilities \citep[see e.g.][]{sampoorna_polarized_2016, hauschildt_improved_2004} in case of larger Doppler shifts. In this paper, we will derive a similar algorithm, and provide a novel, and intuitive method to obtain a numerically stable numerical discretization.
Afterwards, we will evaluate the accuracy and computational efficiency of the new solver by computing the radiation field on an AGB hydrodynamics simulation \citep[taken from][]{malfait_impact_2024} and comparing it to the results using a reference Feautrier solver \citep{feautrier_sur_1964}. For accurate results, we require an appropriate angular discretization for computing the intensity field. We have therefore implemented an adaptive angular discretization scheme, which generates a different set of directions for each model position.

This paper is structured as follows: 
Section \ref{section: comoving solver} contains the derivation of the numerical discretization. Afterwards, we will stabilize the new solver in Section \ref{section: Frequency matching} using an intuitive method and provide a simple illustration why this stabilization is necessary.
Afterwards, we describe our implementation of an adaptive angular discretization scheme in Section \ref{section: adaptive angular discretization}. 
Finally, we apply this new method to a 3D hydrodynamics model in Section \ref{section: application}, and analyze the resulting speedup and accuracy.

\section{Comoving frame solver}\label{section: comoving solver}

Solving the radiative transfer equation gives us the difference in intensity between successive positions. Thus, to calculate the intensity in a specific direction using ray-tracing, the intensity at the previous position in that direction is required. For computational efficiency when computing the radiation field, it would be efficient to use the computed intensity at the previous position both for calculating the mean intensity at that position and to compute the intensity at the next position. However non-zero Doppler shifts change the frequency of the line center in the observer frame.
To address this issue, we start from the comoving frame equation of radiative transfer for a non-relativistic steady flow \citep[see Eq. 19.156 in][]{hubeny_theory_2014}
\begin{equation}\label{eq: comoving equation start}
	\frac{\partial I(x, \nu)}{\partial x} - \frac{\nu}{c}\left(\frac{\partial \mathrm{v}}{\partial x}\right)\frac{\partial I(x, \nu)}{\partial \nu} = \eta(x, \nu) - \chi(x, \nu)I(x, \nu)
\end{equation}
in which $I [\text{W} \text{ sr}^{-1} \text{ m}^{-2} \text{ Hz}^{-1}]$ denotes the specific monochromatic intensity, $x[\text{m}]$ is the position, $\mathrm{v}[\text{m s}^{-1}]$ is the velocity of the medium, $\nu[\text{Hz}]$ is the frequency in the comoving frame, $\eta[\text{W}\text{ sr}^{-1} \text{m}^{-3} \text{ Hz}^{-1}]$ is the total emissivity and $\chi[\text{m}^{-1}]$ is the total opacity.
We then adapt the equation by moving the $\frac{\partial I}{\partial \nu}$ term to the other side, and to improve the numerical stability of the resulting discretization, we add the term $\frac{\partial I}{\partial \nu}\frac{\partial \nu_{\text{extra}}}{\partial x}$ to both sides.
This allows for some freedom in the resulting numerical discretization, as the choice of $\nu_{\text{extra}}(x)$ changes the comoving frame frequency along the path. Note that we have not yet defined $\nu_{\text{extra}}(x)$. For stability reasons (as explained in Section \ref{section: Frequency matching}), this term should be chosen to connect the (yet to be discretized) frequencies at the successive spatial points such that the total frequency change in the observer frame is minimized. The resulting equation becomes:
\begin{equation}\label{eq: comoving equation before characteristics}
\frac{\partial I}{\partial x} + \frac{\partial I}{\partial \nu}\frac{\partial \nu_{\text{extra}}}{\partial x} = \eta(x, \nu) - \chi(x, \nu)I + 
\frac{\partial I}{\partial \nu}\left[\frac{\nu}{c}\left(\frac{\partial \mathrm{v}}{\partial x}\right) + \frac{\partial \nu_{\text{extra}}}{\partial x}\right].
\end{equation}
Assuming a non-relativistic Doppler shift, the term between square brackets is the total frequency change in the observer frame $\frac{\partial \nu_{\text{obs}}}{\partial x}$, to which both the comoving frame and the extra frequency term $\nu_{\text{extra}}$ contribute.

\begin{figure}
	\includegraphics[width=\linewidth]{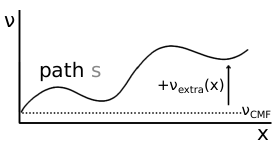}
	\caption{Illustration of how the frequency and position change over the path s, which is used for solving Eq. \ref{eq: comoving equation}. By adding the term $\frac{\partial I}{\partial \nu}\frac{\partial \nu_{\text{extra}}}{\partial x}$ to both sides of Eq. \ref{eq: comoving equation before characteristics}, the comoving frame frequency $\nu$ is different from a constant value $\nu_{\text{CMF}}$ which would be used in Eq. \ref{eq: comoving equation start}.}\label{figure: method of characteristics path illustration}
\end{figure}

By defining a path $s$ such that\footnote{The path defined by the set of equations $x(s) = s, \nu(s) = \nu_{\text{CMF}} + \nu_\text{extra}(x)$, in which $\nu_\text{CMF}$ is a constant, satisfies both equations \ref{eq: s = x + C} and \ref{eq: frequency change}.} 
\begin{align}
\frac{dx}{ds} &= 1\label{eq: s = x + C}\\
\frac{d\nu}{ds} &= \frac{\partial \nu_{\text{extra}}}{\partial x},\label{eq: frequency change}
\end{align}
we can apply a change of variables to the left hand side of Eq. \ref{eq: comoving equation before characteristics} in order to obtain 
\begin{align}
\frac{dI}{ds} &= \eta(x,\nu)-\chi(x,\nu) I(x,\nu) + \frac{\partial I}{\partial \nu}\frac{\partial \nu_{\text{obs}}}{\partial x}\label{eq: comoving equation}.
\end{align}
In this, $s$ is the path length (see Figure \ref{figure: method of characteristics path illustration}). Given that $s$ and $x$ are equivalent (except for an additive constant, see Eq. \ref{eq: s = x + C}), we will replace $s$ with $x$ in any future equation.

By defining the optical depth increment $\Delta\tau_i$ along this path with comoving frame frequency $\nu_i(x)$\footnote{For consistency in notation, we already introduce the frequency index $i$, and the spatial index $j$.}
\begin{equation}\label{eq: moving optical depth}
\Delta\tau_i = \int_{x_j}^{x_{j+1}} \chi(s, \nu_i(s)) ds,
\end{equation}
we solve equation \ref{eq: comoving equation} in a similar manner to the formal solution \citep[see e.g.,][]{kunasz_short_1988}

\begin{equation}\label{eq: formal solution reformulation}
I_{j+1,i}=I_{j,i} e^{-\Delta\tau_i}+e^{-\Delta\tau_i}\int_{0}^{\Delta\tau_i}e^\tau\left(S+\frac{d\nu_{\text{obs},i}}{d\tau}\frac{\partial I}{\partial \nu}\right)d\tau,
\end{equation}
in which we denote the discretized intensity $I(x_j, \nu_i(x_j))$ as $I_{j,i}$ and we define the source function $S [\text{W} \text{ sr}^{-1} \text{ m}^{-2} \text{ Hz}^{-1}]$ to be
\begin{equation}
	S(x, \nu) = \frac{\eta(x,\nu)}{\chi(x,\nu)}.
\end{equation}
Given the similarity in its definition with the observer frame optical depth, we will call $\tau_i$ the effective optical depth.

By discretizing this equation in a similar way to the first order accurate formal solution-type solvers, we end up with the following discretization
\begin{equation}\label{eq: moving solver discretization}
I_{j+1,i}=I_{j,i} e^{-\Delta\tau_i}+\text{Source}+\text{Shift}
\end{equation}
in which the source term is given by

\begin{align}
\text{Source}&=S_{j,i}\left(\frac{1-e^{-\Delta\tau_i}-\Delta\tau_i e^{-\Delta\tau_i}}{\Delta\tau_i}\right)+S_{j+1,i}\left(\frac{\Delta\tau_i -1+e^{-\Delta\tau_i}}{\Delta\tau_i}\right),
\end{align} 
in which $S_{j,i} = S(x_j,\nu_i(x_j))$ is the source function at a specific location and frequency.
The shift term is given by, \footnote{We have replaced the term $\frac{d\nu_{\text{obs}}}{d\tau}$ in equation \ref{eq: formal solution reformulation} with the constant $\frac{\Delta \nu_{\text{obs}}}{\Delta\tau}$. This is an approximation, as we typically assume the frequency to change linearly with the distance $x$ in order to compute the optical depth increment $\Delta\tau$.}

\begin{align}
\text{Shift}=&\frac{\Delta \nu_{\text{obs},i}}{\Delta\tau_i}\frac{\partial I}{\partial \nu}\vert_{x_j,\nu_i(x_j)}\left(\frac{1-e^{-\Delta\tau_i}-\Delta\tau_i e^{-\Delta\tau_i}}{\Delta\tau_i}\right)\\
+&\frac{\Delta \nu_{\text{obs},i}}{\Delta\tau_i}\frac{\partial I}{\partial \nu}\vert_{x_{j+1},\nu_i(x_{j+1})}\left(\frac{\Delta\tau_i -1+e^{-\Delta\tau_i}}{\Delta\tau_i}\right)\nonumber
\end{align}

in which $\Delta \nu_{\text{obs},i}$ denotes the frequency difference in the observer frame and the subscript denotes at which point to evaluate the source function and derivative. We denote with $\nu_{\text{obs,}i}(x)$ the required frequencies in the observer frame used at each position to discretize this equation. Therefore, $\Delta\nu_{\text{obs},i}=\nu_{\text{obs,}i}(x_{j+1})-\nu_{\text{obs,}i}(x_j)$.

To complete the discretization, we utilize a second order accurate forward discretization for the derivative of the intensity with respect to the frequency $\frac{\partial I}{\partial \nu}$
\begin{align}
\frac{\partial I}{\partial \nu}\vert_{x=x_j,\nu=\nu_i(x_j)}\simeq \sum_{k = 0}^2 c_{ijk} I(x_j,\nu_{i+k}(x_j))
\end{align}
for which the coefficients $c_{ijk}$ are derived in \ref{appendix: coefficients second order freq derivative} and in which $\nu_i(x_j)$ are the discretized frequencies in the comoving frame at each position.

To efficiently solve the coupled set of Equations \ref{eq: moving solver discretization} for a single spatial increment, we will impose one assumption. We will demand that the stencil is chosen such that all frequency differences $\Delta\nu_{\text{obs},i}$ in the observer frame have the same sign, which is given by sign of the Doppler shift. After applying appropriate boundary conditions, the resulting set of equations will become a matrix equation with an upper or lower diagonal matrix on the left hand side, similar in shape to the left hand side of Eq. \ref{eq: comoving stability matrix equation}. We solve this set of equations using Gaussian elimination in $O(N_\text{frequencies})$ time.

\subsection{How to practically compute the effective optical depth}
In Eq. \ref{eq: moving optical depth}, we have defined the optical depth increments $\Delta\tau_i$ using a non-constant frequency path $\nu_i(x)$. We now explain how to evaluate this expression, as this formula looks similar to the observer frame optical depth, but is slightly different.

Starting from any computation method for the optical depth, when filling in any quantity at a given position $x$, we also need to substitute the corresponding frequency $\nu(x)$ at that position. For example, in case of the trapezoidal rule, the optical depth increment $\Delta\tau_i$ becomes
\begin{align}
\Delta\tau_i = \Delta x\frac{\chi(x_j, \nu_i(x_j))+\chi(x_{j+1}, \nu_i(x_{j+1}))}{2},
\end{align}
in which $\Delta x$ is the distance increment.

\subsection{Instabilities}\label{section: instabilities}
In two papers \citep{sampoorna_polarized_2016, hauschildt_improved_2004}, numerical instabilities have been observed when applying the method of \cite{baron_co-moving_2004} to models with large velocity gradients, but the origin of the instabilities was not discussed in these papers. We argue that the numerical instabilities are caused by an extrapolation in the frequency discretization, based on the stability criterion we derive in Section \ref{section: stability analysis}.
The criterion states that numerical stability becomes worse as the Doppler shift increases relative to the spacing between adjacent frequencies in the frequency discretization \citep[see also Section 6.6 in][]{castor_radiation_2004}.

\begin{figure}
	\centering
	\includegraphics[width=\linewidth]{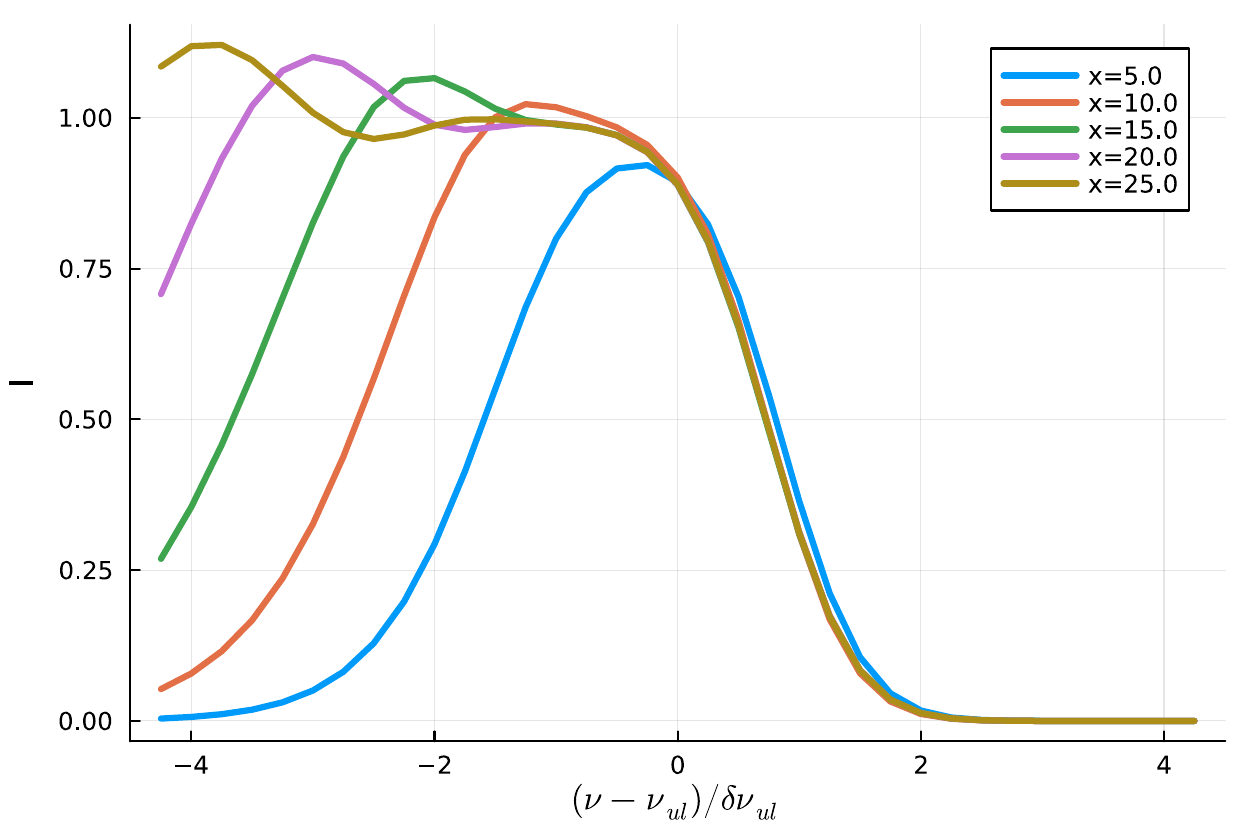}
	\caption{Illustration of the numerical behavior of the comoving solver, when following the Doppler shift in order to determine the discretization. All intensity profiles at the different positions are plotted in the comoving frame, centered around the line center $\nu_{ul}$. Numerical oscillations are present in the computed intensity.}\label{figure: julia example oscillatory instability}
\end{figure}

To illustrate the instabilities which can occur, we have created a simple 1D single ray model with a single line transition $u\to l$ using a Gaussian line profile $\phi_{ul}$
\begin{align}\label{eq: gaussian}
\phi_{ul}(x, \nu) = \frac{1}{\delta\nu_{ul}(x)\sqrt{\pi}}e^{-\frac{(\nu-\nu_{ul})^2}{\delta\nu_{ul}(x)^2}}\quad,
\end{align}
in which $\delta\nu_{ul}$ is the line width and $\nu_{ul}$ is the line center.
The setup is described in Table \ref{table: julia example setup}. To properly showcase the possible numerical instabilities, we have defined this model with high optical depths and moderate Doppler shift. As we have defined the model such that the source function $S=1.0$ is constant, and the initial intensity is $0$, we expect from the formal solution that the intensity $I$ in the entire frequency range is bounded from above by $1.0$. However, when applying the comoving solver, if the path $s$ is following the Doppler shift in frequency space, we observe oscillations (see Figure \ref{figure: julia example oscillatory instability}), pushing the computed intensities above the value of the source function, which is nonphysical.

\begin{table}
	\caption{Parameters for the 1D model with a constant velocity gradient. To evaluate the intensities, we use a uniform frequency discretization with spacing $\Delta\nu_s$. Between successive points at distance $\Delta x$, the Doppler shift causes a constant observer frame frequency difference $\Delta v_{\text{obs}}$. The boundary intensity is set to $0$. 
	All units are re-scaled in dimensionless units such that the line width $\delta\nu_{ul} = 1.0$ and the line optical depth $\int_0^1 \chi/\phi_{ul}\enspace dx$ over a unit distance is $1.0$.
	}\label{table: julia example setup}
	\centering
	\begin{tabular}{c|c|c|c|c|c}
		$\chi/\phi_{ul}$ & $\eta/\phi_{ul}$ & S & $\Delta\nu_s$ & $\Delta x$ & $\Delta v_{\text{obs}}$\\
		\hline
		$1.0$&$1.0$&$1.0$&$\delta\nu_{ul}/4$&$5.0$&$1.1\delta\nu_{ul}$		
	\end{tabular}
\end{table}

\subsection{Derivation of the stability of the comoving solver}\label{section: stability analysis}

Given the numerical instabilities described in the previous section, we derive the stability criterion for our discretization in this section. Therefore, 
we apply perturbation analysis to the discretized equations for the position interval $[x_0, x_1]$. The resulting set of equations can be written as a matrix equation, for which we will compute the eigenvalues. We have assumed the frequency differences $\Delta\nu_i$ to be positive (a similar derivation can be made for negative $\Delta \nu_i$).
To complete the set of equations, boundary conditions have been posed on $I(x_1,\nu_{n-1}(x_1)), I(x_1,\nu_{n}(x_1))$.

\begin{align}\label{eq: comoving stability matrix equation}
&\begin{bmatrix}
1+A_1&B_1&C_1&\dots&0\\
\vdots&\ddots&\ddots&\ddots&\vdots\\
0&\dots&1+A_{n-2}&B_{n-2}& C_{n-2}\\
0&\dots&0&1&0\\
0&\dots&0&0&1
\end{bmatrix}
\begin{bmatrix}
\delta I_{1,1}\\
\vdots\\
\delta I_{1,n-2}\\
\delta I_{1,n-1}\\
\delta I_{1,n}\\
\end{bmatrix}
=\nonumber\\	
&\begin{bmatrix}
e^{-\Delta\tau_1}+A_1'&B_1'&C_1'&\dots&0\\
\vdots&\ddots&\ddots&\ddots&\vdots\\
0&\dots&e^{-\Delta\tau_{n-2}}+A_{n-2}'&B_{n-2}'& C_{n-2}'\\
0&\dots&0&0&0\\
0&\dots&0&0&0
\end{bmatrix}
\begin{bmatrix}
\delta I_{0,1}\\
\vdots\\
\delta I_{0,n-2}\\
\delta I_{0,n-1}\\
\delta I_{0,n}\\
\end{bmatrix},
\end{align}
in which we define $\delta I_{j,i}$ as the perturbation on $I(x_j, \nu_i(x_j))$. In this equation,
\begin{align*}
	A_i&=-c_{i10}\frac{\Delta \nu_{\text{obs,}i}}{\Delta\tau_i}\left(\frac{\Delta \tau_i -1+e^{-\Delta\tau_i}}{\Delta\tau_i}\right)\\
	B_i&=-c_{i11}\frac{\Delta \nu_{\text{obs,}i}}{\Delta\tau_i}\left(\frac{\Delta \tau_i -1+e^{-\Delta\tau_i}}{\Delta\tau_i}\right)\\
	C_i&=-c_{i12}\frac{\Delta \nu_{\text{obs,}i}}{\Delta\tau_i}\left(\frac{\Delta \tau_i -1+e^{-\Delta\tau_i}}{\Delta\tau_i}\right)
\end{align*}
On the right-hand side, we have similarly defined 
\begin{align*}
A_i'&=c_{i00}\frac{\Delta \nu_{\text{obs,}i}}{\Delta\tau_i}\left(\frac{1-e^{-\Delta\tau_i}-\Delta\tau_i e^{-\Delta\tau_i}}{\Delta\tau_i}\right)\\
B_i'&=c_{i01}\frac{\Delta \nu_{\text{obs,}i}}{\Delta\tau_i}\left(\frac{1-e^{-\Delta\tau_i}-\Delta\tau_i e^{-\Delta\tau_i}}{\Delta\tau_i}\right)\\
C_i'&=c_{i02}\frac{\Delta \nu_{\text{obs,}i}}{\Delta\tau_i}\left(\frac{1-e^{-\Delta\tau_i}-\Delta\tau_i e^{-\Delta\tau_i}}{\Delta\tau_i}\right)
\end{align*}

The eigenvalues of this system of equations determine the stability of this discretization.
To find them, one can exploit the properties of upper diagonal matrices: the inverse of an upper diagonal matrix also being an upper diagonal matrix and the multiplication of two upper diagonal matrices still being an upper diagonal matrix. Thus, by inverting the matrix on the left-hand side (multiplying it together with the one on the right-hand side), another upper diagonal matrix is obtained. For an upper diagonal matrix, one can simply find the eigenvalues $\lambda_i$ by looking at the diagonal. The resulting diagonal elements are given by: 
\begin{equation}\label{eq: basic stability equation}
\lambda_i=\frac{e^{-\Delta\tau_i}+A_i'}{1+A_i}.
\end{equation}
As equation (\ref{eq: basic stability equation}) is too dense, we will first define the constants $d_i=c_{i10}\Delta\nu_{\text{obs,}i}$, $d_i'=c_{i00}\Delta\nu_{\text{obs,}i}$. These quantities are related to the ratio of frequency difference $\Delta\nu_{\text{obs,}i}$ versus the width of the frequency discretization. Then by rewriting, one finds the stability condition to be
\begin{equation}\label{eq: moving stability condition}
\lambda_i=\frac{\Delta \tau_i^2 e^{-\Delta\tau_i}+d_i'\left(1-e^{-\Delta\tau_i}-\Delta\tau_i e^{-\Delta\tau_i}\right)}{\Delta\tau_i^2-d_i\left(\Delta\tau_i-1+e^{-\Delta\tau_i}\right)}.
\end{equation}
The result is that this discretization is almost unconditionally stable\footnote{Due to the possible difference between $c_{i10}$ and $c_{i00}$, no guarantee exists for stability. This corresponds with the contraction/expansion of the frequency discretization. The existence of maser regimes, for which $\Delta\tau<0$, is also ignored, as the frequency derivative will not be sampled well in that regime.}, meaning that $|\lambda_i|\leq1$, given that coefficients $c_{i00}, c_{i10}<0$ if all frequencies are ordered (see \ref{appendix: coefficients second order freq derivative}). However, oscillatory behavior ($-1\lessapprox\lambda_i<0$) can exist if the parameters $d_i$, $d_i'$ are large. This corresponds to extrapolating the frequency derivative terms.

\section{Frequency matching}\label{section: Frequency matching}
In the preceding section, we have derived the comoving solver and its properties regarding numerical stability. Given the potential for numerical instabilities to produce nonphysical results, we require a method to mitigate these numerical effects without significantly affecting computational efficiency. In this section, we describe a novel, yet intuitive method for stabilizing numerical discretizations of transport equations in order to reduce numerical oscillations in the computed result.

\subsection{A novel way to sidestep numerical instabilities}\label{section: frequency matching explanation}
Upon examining the stability criterion (Eq. \ref{eq: moving stability condition}), two apparent methods emerge to improve numerical stability: either insert extra points in-between for interpolating the velocity difference, therefore reducing the Doppler shifts between successive spatial positions, or coarsen the frequency discretization. However, neither solution is appropriate for our purposes: in the former case, the computation time increases, contradicting the objective of this paper, while in the latter case, the accuracy decreases when computing the mean line intensity. In order to reach our goal of numerical stability, we must explore alternative approaches.

\begin{figure}
	\centering
	\includegraphics[width=\linewidth]{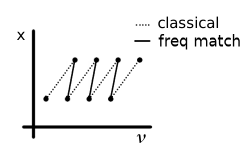}
	\caption{Illustration of the different ways to match frequencies in the comoving solver. 
	Intuitively, one would connect the frequencies by following the doppler shift (classical). By minimizing the frequency difference in the observer frame instead (freq match), numerical instabilities can be suppressed, at the cost of extra boundary conditions, as less frequency points are connected.
	}\label{figure: freq match illustration}
\end{figure}

\begin{figure}
	\centering
	\includegraphics[width=\linewidth]{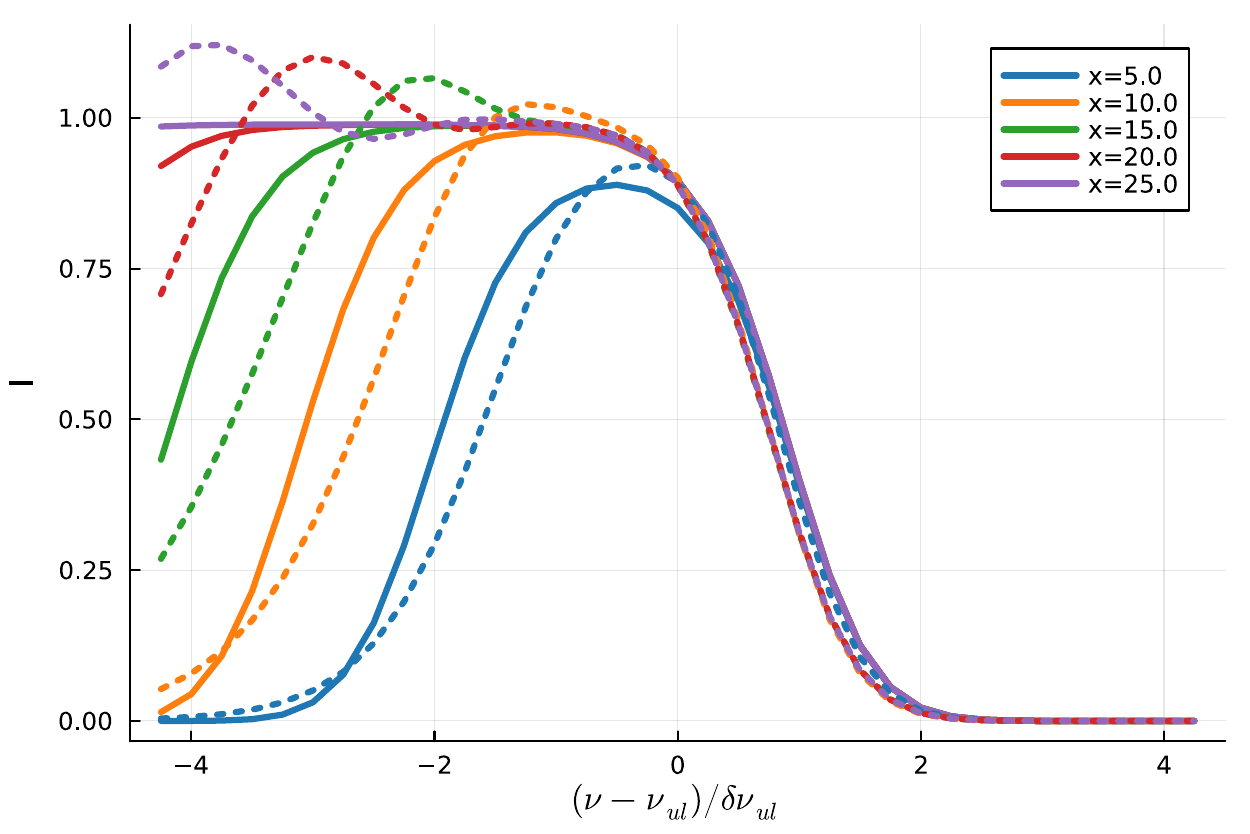}
	\caption{Illustration of the numerical behavior of the comoving solver, when matching the frequencies in order to determine the discretization. No numerical oscillations can be seen in this plot, in contrast to when letting the frequencies follow the Doppler shift. For a better comparison with Figure \ref{figure: julia example oscillatory instability}, we have added the data of that figure as dotted lines in this figure.}\label{figure: julia example freq match}
\end{figure}

The main stability issue arises from the extrapolation of the frequency differences, due to the change in frequency being dictated by the Doppler shift. Therefore, the stability issues could be resolved by minimizing the frequency changes. This can be achieved by altering how to connect the frequency discretizations of successive spatial positions. 
The most straightforward approach involves mapping all frequency indices one-to-one from low to high frequency. However, mapping all frequency indices is not required. Instead, we can do the mapping in such a way which ensures the frequency differences to be minimized. This solves our numerical stability issues at the cost of extra boundary conditions (see Figure \ref{figure: freq match illustration} and Sect. \ref{section: boundary conditions}). To illustrate the impact of this method, we apply it to the simple 1D ray model described in Table \ref{table: julia example setup}. By mapping the frequency discretization such that the observer frame frequency differences $\Delta\nu_{\text{obs},i}$ are minimal, we obtain the results in Figure \ref{figure: julia example freq match}. No numerical oscillations occur, and the computed intensities remain below the value of the constant source function $S$.

For practically implementing this method, we recommend using the observer frame to evaluate the frequency difference, as the objective is to minimize $\left|\Delta\nu_{\text{obs},i}\right|$ when constructing the stencil. Assuming frequency discretizations $\nu_{\text{obs},i}(x_j), \nu_{\text{obs},i}(x_{j+1})$, determine the stencil for all $\nu_{\text{obs,}i}(x_{j+1})$ such that 
\begin{align}
	\left|\Delta\nu_{\text{obs},i}\right| = \left|\nu_{\text{obs},i_{j+1}}(x_{j+1})-\nu_{\text{obs},i_{j}}(x_j)\right|
\end{align} is minimized, in which $i_{j}, i_{j+1}$ are the frequency indices to connnect. 
For ease of solving the equations, make sure that the sign of $\nu_{\text{obs,}i_{j+1}}(x_{j+1})-\nu_{\text{obs,}i_{j}}(x_{j})$ is positive for all $\nu_{\text{obs,}i}(x_{j+1})$ in case of a positive Doppler shift ($\Delta\mathrm{v}>0$) and vice versa. In this way, we either obtain an upper or a lower diagonal matrix for the system of equations to solve.
In Section \ref{section: boundary conditions}, we will explain how to implement the boundary conditions.

\subsection{Dealing with boundary conditions}\label{section: boundary conditions}
Matching the frequencies introduces extra boundary conditions, due to not mapping the entire frequency discretization. Given the complexity of implementing these boundary conditions, we provide a brief overview of our approach in this section. A more comprehensive explanation can be found in \ref{Appendix: boundary condition implementation}.

\begin{figure}
	\includegraphics[width=\linewidth]{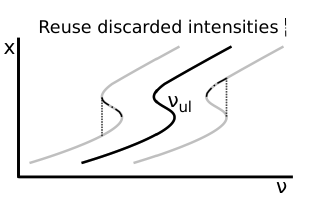}
	\caption{Illustration of why extra boundary conditions are required for the comoving solver. The line center $\nu_{ul}$, together with the bounds of the frequency discretization (in grey) move according to the Doppler shift. As we shift the discretization, the computed intensities at frequencies outside of the current frequency discretization will be discarded. Inside the dash-dotted regions, we require previously computed intensity information as we re-encounter these frequencies.
	}\label{figure: boundary conditions}
\end{figure}

In general hydrodynamics models, the velocity profile is usually non-monotonic. In this paper, we assume the intensities to only change due to the influence of molecular lines, meaning that the intensities at frequencies sufficiently far enough from the line centers, remain constant. The comoving method discards information at the frequencies it is moving away from. Consequently, during any later computation, it might re-encounter frequencies for which it has discarded the previously computed intensities, as illustrated in Fig. \ref{figure: boundary conditions}. To address this issue, we have implemented two different approaches. A first approach involves temporarily storing the computed intensities and reusing them when re-encountering the corresponding frequencies, which requires extensive bookkeeping.
Alternatively, for a simpler, more approximate implementation, one can disregard the previously computed intensities entirely, always filling in the boundary conditions using black-body intensities. However, this method is only strictly valid if the non-monotonicity of the velocity field is minor compared to the total width of the frequency discretization. While we expect the former approach to be more accurate, it is also more time-consuming than the latter approach.\\

\subsection{A benchmark with non-monotonic velocity field}\label{section: comoving benchmark}

\begin{figure}
	\includegraphics[width=\linewidth]{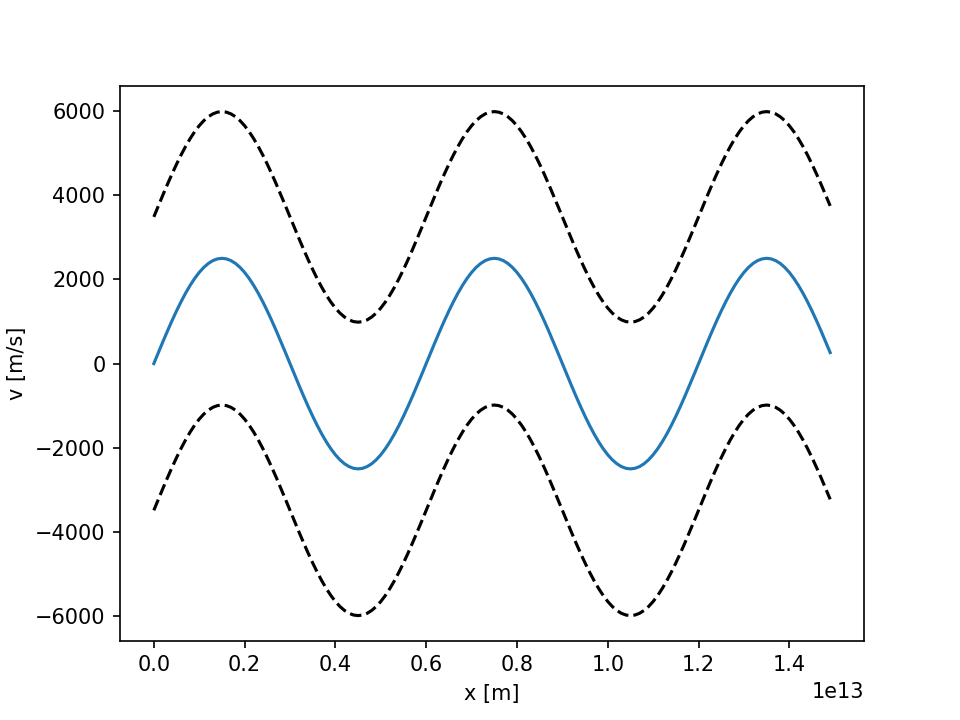}
	\caption{The non-monotonic velocity profile of the comoving solver boundary condition benchmark. Both the velocity at each position and corresponding bounds of the frequency discretization (converted to velocity using the Doppler shift) are present. For most of the domain, the boundary conditions we evaluate require previously computed intensities (see Fig. \ref{figure: boundary conditions}).}
	\label{figure: comoving solver benchmark velocity}
\end{figure}

To show the impact of the different treatments of the frequency boundary condition (see Section \ref{section: boundary conditions}) on the accuracy, we devise a minimal benchmark with a non-monotonic velocity field to compare the computed mean line intensities $J_{ul}$ (see Eq. \ref{eq: mean line intensity}) of the different solvers. This benchmark involves a 1D model of a single ray, where we compute intensities for the CO v$=0$ J=$1-0$ line\footnote{The line data is obtained from the \textsc{LAMDA} database, available at \url{https://home.strw.leidenuniv.nl/~moldata/}.}.
We will compare the mean line intensity computed using the comoving solver against the value computed using an established solver, the Feautrier solver \citep{feautrier_sur_1964}, as implemented in \textsc{Magritte} \citep[see e.g.][]{ceulemans_magritte_2024, de_ceuster_3d_2022}. We will test both implementations of the frequency boundary conditions. To properly assess the capability of the solver in handling non-monotonic velocity fields, we opt for a sinusoidal velocity prescription (see Fig. \ref{figure: comoving solver benchmark velocity}), with its amplitude several times the width of the frequency discretization around each line. Since this is a benchmark for the non-monotonic behavior of the velocity field, we keep all other parameters constant. The velocity profile is given by a sinusoidal function with period of $60$ model points and amplitude $2.5\cdot 10^3 \text{ m/s}$. The model assumes LTE, a Gaussian line profile (see Eq. \ref{eq: gaussian})
and corresponding line width $\delta\nu_{ul}$ computed according to
\begin{align}\label{eq: line width}
\delta\nu_{ul}(x) = \frac{\nu_{ul}}{c}\sqrt{\frac{2k_bT}{m_{\text{spec}}} + v_{\text{turb}}^2}.
\end{align}
In this equation, $k_b = 1.38 \cdot 10^{-23} \text{J} \text{ K}^{-1}$ denotes Boltzmann's constant, $T [\text{K}]$ represents the local gas temperature, $m_{\text{spec}} [\text{kg}]$ is the gas species mass and $v_{\text{turb}} [\text{m}\text{ s}^{-1}]$ is the turbulent velocity.

We use a Gauss-Hermite quadrature, using $25$ frequencies for the frequency discretization.
The spatial domain is linearly discretized with $150$ points, using a spacing of $1.0\cdot 10^{11} \text{ m}$ between them. Finally, the gas temperature is $500 \text{ K}$ and the CO number density is given by $1.0\cdot 10^{8} \text{ m}^{-3}$.
This benchmark represents a simplified version of a slice of the spiral outflow originating from the binary interaction between an AGB star and a companion, which represents our application in Section \ref{section: application}.

In this benchmark, a correct treatment of the frequency boundary is required (compare Fig. \ref{figure: comoving solver benchmark velocity} with Fig. \ref{figure: boundary conditions}). We will compare the results of the different solvers using the mean line intensity 
\begin{equation}\label{eq: mean line intensity}
J_{ul}(\boldsymbol{x})=\frac{1}{4\pi}\oint_{\Omega}\int_0^{+\infty} I(\hat{\boldsymbol{n}},\boldsymbol{x},\nu)\phi_{ul}(\nu)\text{d}\Omega \text{d}\nu.
\end{equation}

\begin{figure}
	\includegraphics[width=\linewidth]{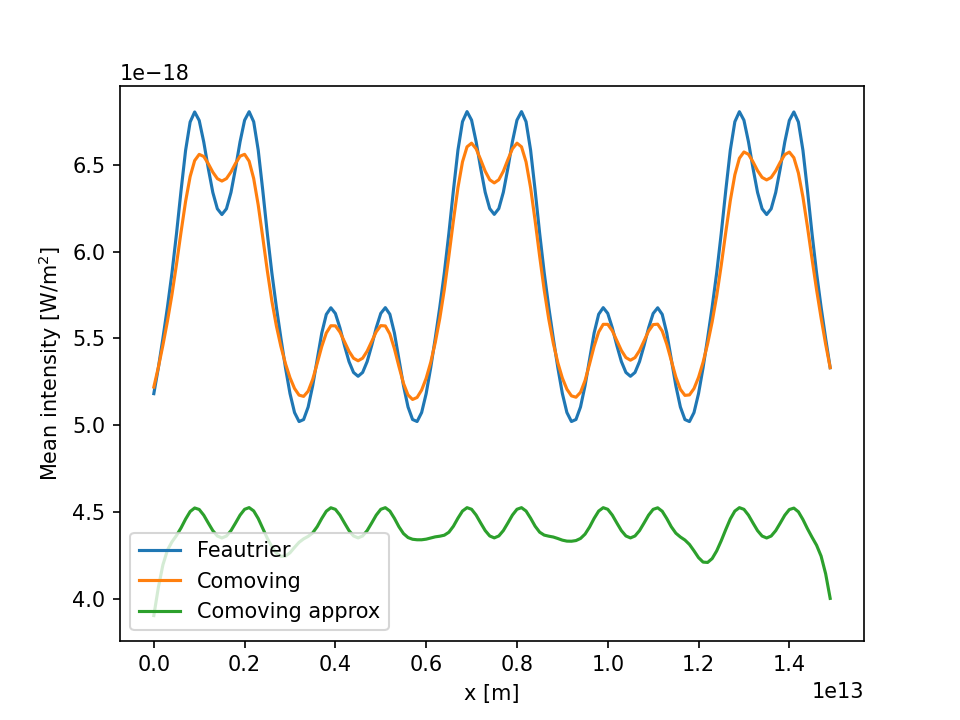}
	\caption{The computed mean line intensity $J_{ul}$ for the CO J=1-0 v=0 line, for the 1D benchmark model with a sinusoidal velocity field. The computed results from the comoving solver with correct frequency boundary conditions agree with the reference values obtained from the Feautrier solver. The comoving solver with approximate boundary conditions fails the benchmark, given that the velocity field of the model is significantly non-monotonic.}
	\label{figure: comoving solver benchmark intensity}
\end{figure}

As expected, the approximate version of the comoving solver performs poorly (see Fig. \ref{figure: comoving solver benchmark intensity}). The computed mean line intensity is significantly lower than the reference intensity because this implementation does not store the previously computed intensities. The version with correct boundary conditions obtains similar results to the reference Feautrier solver, with a mean absolute relative difference in mean line intensity $J_{ul}$ of $2\%$ for the CO v=0 J=1-0 line. However, the comoving solver is more dissipative than the Feautrier solver.
This benchmark served a proof of concept, to show that we can apply the comoving solver to non-monotonic velocity field. We refer the reader to Section \ref{section: application} for an application of this solver to a 3D hydrodynamics model. Please note that the result of this benchmark does not imply that one must always use fully correct frequency boundary conditions. In situations where the non-monotonicity of the velocity field is insignificant, using approximate frequency boundary conditions, which require less computation time, may suffice.\\

\section{Adaptive angular discretization}\label{section: adaptive angular discretization}

When computing angle-averaged mean intensities in a model using ray-tracing, the chosen angular discretization impacts the results. The simplest option is to use a uniform angular discretization at each position in the model.
However, when the incoming intensity varies significantly depending on the angle, this approach can result in numerical artifacts, as smaller angular features cannot be resolved with a coarse discretization. For example, when applying radiative transfer on the hydrodynamics model in Section \ref{section: application}, the computed mean intensity field shows rays originating from the dense model center, due to the higher intensity originating from this small, dense region (see Fig. \ref{figure: intensity static vs adaptive angular discretization}).

To solve this numerical issue, we will not naively increase the density of the angular discretization, as this will significantly increase the overall computational cost. Instead, we will determine for each position a good set of directions to use. For every point, we estimate the density of the direction vectors towards the other points in the model on a fine discretization of the entire solid angle. The regions of the solid angle with the highest density require a fine angular resolution, while the other regions can be approximated using coarse angular resolution.

\subsection{Adaptive angular discretization algorithm}\label{section: adaptive directions algorithm}
In this section, we will explain the algorithm \citep[inspired by][]{deceuster_magritte_2020} that we use to generate the adaptive angular discretization, and how to apply this to the comoving solver.


We initialize the algorithm by specifying the discretization level of the \textsc{Healpix} sphere. In this paper, we have chosen the 5th discretization level, which contains to $12\cdot2^{2\cdot5} = 12288$ pixels. Since the computation time for the mean intensity scales linearly with the number of angles, we want to minimize the number of directions. Every individual point requires a different angular discretization, as the regions of interest lie in a different position relatively to the position of the point. For computation time scaling purposes\footnote{Figuring out at each position in which pixel all other points lie, costs $O(N_{\text{points}}^2)$ computation time. A randomly chosen subset will be used as a proxy for the location of all points.}, we will first select a randomly chosen set of points used to infer which model regions should be sampled more densely (totaling 10000 points in this paper). To shorten notation, we define the ``pixel count'' to be the number of sample points for which their relative position to the current point has a direction contained in the pixel of the \textsc{Healpix} \citep{gorski_healpix_2005} discretization of the sphere. We give the algorithm for generating the adaptive angular discretization for a single position in Listing \ref{listing: flowchart}.

\begin{lstlisting}[caption={Adaptive direction discretization algorithm. We denote the pixel count with $P(a, n)$, using angle index a, discretization level $n\in 0,\dots, N$, with $N$ being the finest level. The \textsc{Healpix} angular discretization recursively subdivides, such that $P(a,N-1)=\sum_{i = 0}^3 P(4a+i, N)$. Finally, $M$ is the maximum amount of pixels to subdivide per discretization level ($2$ in this paper).}, numbers=left, mathescape, label={listing: flowchart}, numbers=left, breaklines=true, basicstyle=\small, escapechar=@, breakautoindent=false, breakindent=2ex]
@Create empty list $\mathcal{L}$ of discretized angles@
@Compute pixel count $P(a, N)$@
@Compute $P(a, N-1)$ = $\sum_{i = 0}^3 P(4a+i, N)$@
@Subdivide M pixels with highest $P(a, N-1)$, adding 4M pixels to $\mathcal{L}$@
@Compute $P(a, N-2)$ = $\sum_{i = 0}^3 P(4a+i, N-1)$\label{listingline:startloop}@
@Identify pixels at level $N-2$ which contain pixels in $\mathcal{L}$, denote count with $C\leq M$@
@Add the corresponding subpixels which do not yet overlap with any pixels in $\mathcal{L}$, to $\mathcal{L}$@
@Find other M-C pixels with highest $P(a, N-2)$, adding 4(M-C) pixels to $\mathcal{L}$\label{listingline:endloop}@
@Repeat steps \ref*{listingline:startloop} to \ref*{listingline:endloop} for each coarser level until either the entire solid angle is subdivided or the coarsest level is reached. Add remaining pixels at level 0 to $\mathcal{L}$, such that $\mathcal{L}$ covers the entire solid angle.@
@Return $\mathcal{L}$@
\end{lstlisting}

Note that as \textsc{Magritte} requires point-wise symmetric directions (needed for the Feautrier solver), we apply this algorithm to only half of the solid angle. To achieve this, we have mapped the entire sphere onto half a sphere by considering the maximal count of the point-wise symmetric pixels over both halves.

The algorithm above describes the angular discretization used in the Feautrier solver. For the comoving solver, we need to slightly modify the ray-tracing algorithms. As a first implementation, we will only reuse intensities if the direction of the ray is a part of the angular discretization of another point on the ray. Do note that this results in a reduced efficiency of the comoving solver when compared to using a uniform angular discretization, as less information will be reused, and therefore radiative transfer will need to be computed on more rays. As future work, an interpolation scheme might be implemented in order to evaluate the mean intensity using intensities from directions which do not exactly match the angular discretization at a specific point. However, this is out of scope for this paper.\\

\section{Application}\label{section: application}
In the previous sections, we considered 1D radiative transfer. For quantifying the speed and accuracy of the new solver, we have implemented the comoving solver in \textsc{Magritte} \citep[][]{ceulemans_magritte_2024, de_ceuster_3d_2022, deceuster_magritte_2020, deceuster_magritte_2019} and we will compare it against the well-established Feautrier solver \citep{feautrier_sur_1964} by computing NLTE line radiative transfer in a 3D Phantom \citep{price_phantom_2018} model. As \textsc{Magritte} is a ray-based 3D radiative transfer code, it computes the radiation field by solving many 1D radiative transfer problems\footnote{For the radiation field, we require the intensity at every position, discretized direction and discretized frequency in the comoving frame.}. The comoving solver can therefore replace the Feautrier solver in the intensity calculation step.

The model ``v10e00'' taken from \cite{malfait_impact_2024} is a smoothed particle hydrodynamics simulation of an AGB star with a companion star. The model parameters can be found in Table \ref{table: phantom model parameters}. In this model, we assume a CO/H$_2$ abundance ratio of $10^{-4}$ and will be using the CO v=0 J=0 to J=40 v=0 levels available from the \textsc{LAMDA} database \citep{schoier_atomic_2005}. Velocities, temperatures and H$_2$ densities are taken from the hydrodynamics model. We have implemented two versions of the comoving solver, each using a different treatment of the boundary conditions, as explained in Section \ref{section: boundary conditions}. The version with approximate boundary condition will be denoted by `comoving approx'. Since the timings and accuracy results between these two approaches vary, we will test both on the model to compare them versus the Feautrier solver. All simulations in this section were run on a server with dual AMD EPYC 7643 socket, utilizing all 192 threads.
We compare the results of the different solvers using the mean line intensity $J_{ul}$ (see Eq. \ref{eq: mean line intensity}).

\begin{table}
	\centering
	\caption{Phantom model ``v10e00" model parameters}\label{table: phantom model parameters}
	\begin{tabular}{|l|r|}
		\hline
		AGB stellar mass $M_p$ & $1.5 \text{M}_\odot$\\
		AGB stellar temperature $T_p$ & $3000$ K\\
		Companion mass $M_s$ & $1 \text{M}_\odot$\\
		Initial wind speed $v_{ini}$ & $10$ km$/$s\\\hline
		Semi-major axis $a$ & $6$ AU\\
		\hline
	\end{tabular}
\end{table}

\subsection{Uniform angular discretization}\label{section: static angular discretization application}
In this section, we compare the solvers using a uniform angular discretization, based on the second discretization level of the Healpix \citep{gorski_healpix_2005}, containing 48 discrete ray directions.

\begin{figure}
	\centering
	\includegraphics[width=\linewidth]{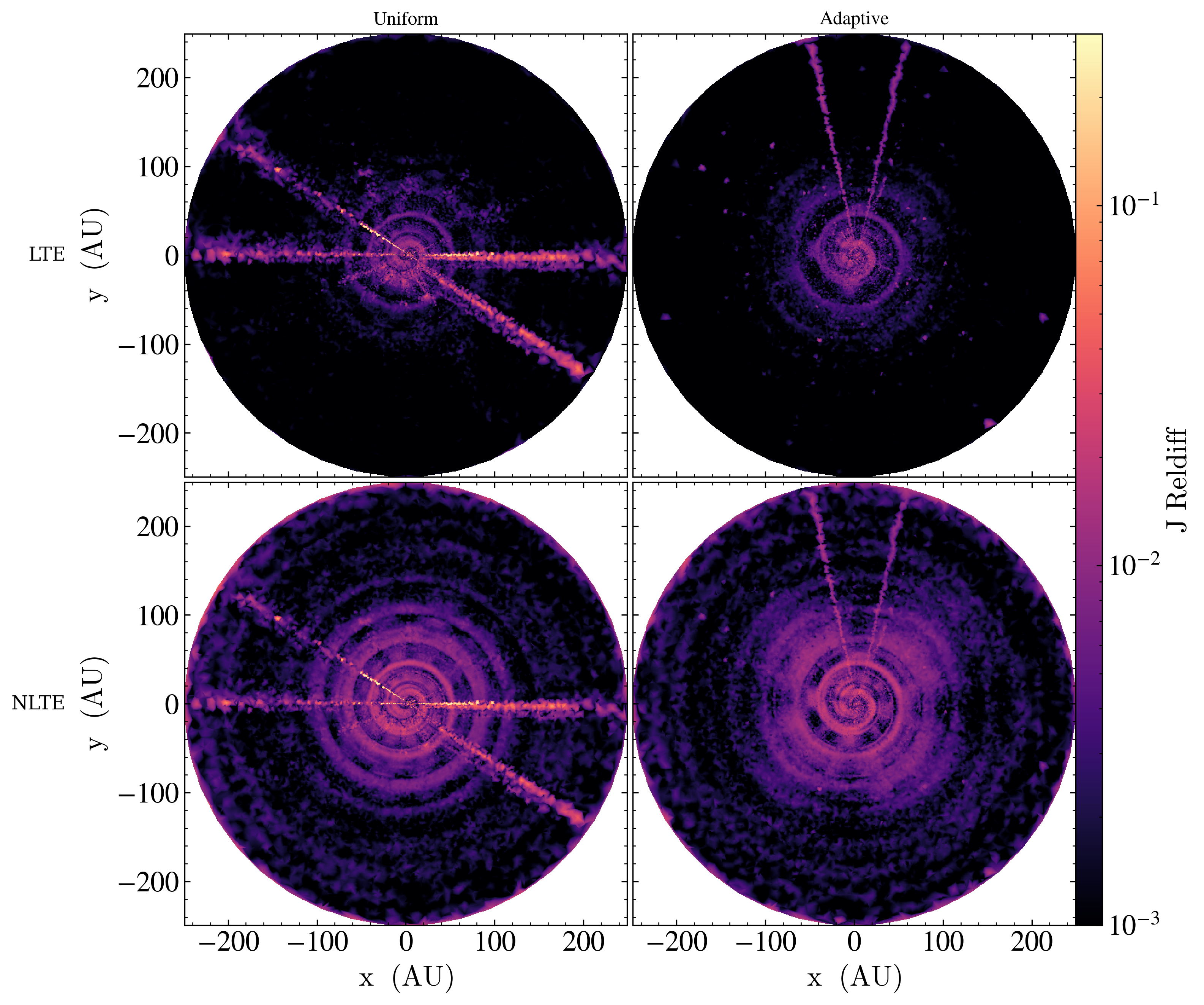}
	\caption{Slice plot of the relative differences between the computed mean line intensity $J_{ul}$ for the CO v=0 J=2-1 line using the Feautrier solver and the comoving solver. Left: using the uniform angular discretization. Right: using the adaptive angular discretization. We plot both the intensity in case of LTE and the self-consistent intensity after convergence (NLTE). On the left, we see high relative differences on straight rays emerging from the model center, which is an effect of the limited angular discretization.}\label{figure: intensity static vs adaptive angular discretization}
\end{figure}

On Fig. \ref{figure: intensity static vs adaptive angular discretization}, we notice that the relative differences between the intensities obtained using the comoving and the Feautrier solvers seem to be quite high on some straight rays which originate from the dense model center. These numerical artifacts have already been explained in Section \ref{section: adaptive angular discretization}. Due to the presence of the aforementioned numerical artifacts, the obtained results might not be fully representative, but for completeness, we still include the timings and accuracy results in this section.
 
\begin{figure}
	\includegraphics[width=\linewidth]{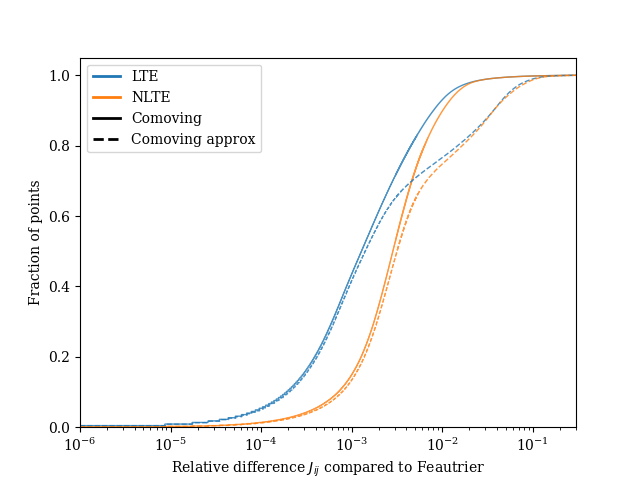}
	\caption{Cumulative difference between the computed mean line intensity $J_{ul}$ for the CO J=2-1 v=0 line using the Feautrier solver and the comoving solvers, when using a uniform angular discretization. Both the differences in case of a single iteration after setting the level populations $n_i$ to LTE (in blue) and the difference after letting the level populations converge (in orange). The figure shows the results for the correct boundary conditions described in Section \ref{section: boundary conditions} as solid lines and the results when using the approximate boundary conditions as dashed lines.}\label{figure: comparison differences comoving implementations static rays}
\end{figure}

We find that the comoving solver with accurate boundary conditions has relative differences in NLTE intensities less than $1\%$ for $89\%$ of model points, when compared to the Feautrier solver (see  Fig. \ref{figure: comparison differences comoving implementations static rays}). The comoving solver with approximate boundary conditions has relative differences in mean intensities below $1\%$ for $75\%$ of the model points. The timings show (see Table \ref{table: timings NLTE static angular discretization}) that both comoving solvers provide significant speedups when compared to the Feautrier solver. The reuse of computed intensities in the comoving solvers results in significantly less rays being used for computing the intensity field, when compared to the Feautrier solver.

\begin{table}
	\centering
	\caption{Time required to compute NLTE level populations (5 iterations) on the \textsc{Phantom} model, containing $885501$ points and using a uniform angular discretization. To illustrate the maximum possible obtainable speedup (assuming all algorithms taking the same amount of time per computed ray), we also add the number of rays used to compute the intensity field per solver. This table compares the Feautrier solver, the comoving solver with correct boundary conditions, and the comoving solver with approximate boundary conditions.}
	\label{table: timings NLTE static angular discretization}
	\begin{tabular}{l|r|r}
		Method&Time[s]&Rays [.]\\
		\hline
		Feautrier& $9.7\cdot 10^3$&$2.1\cdot 10^7$\\
		Comoving & $4.8 \cdot 10^3$&$9.3\cdot 10^5$\\
		Comoving approx& $1.9 \cdot 10^3$&$9.3\cdot 10^5$\\
	\end{tabular}
\end{table}

\subsection{Adaptive angular discretization}\label{section: adaptive angular discretization application}

\begin{figure}
\includegraphics[width=\linewidth]{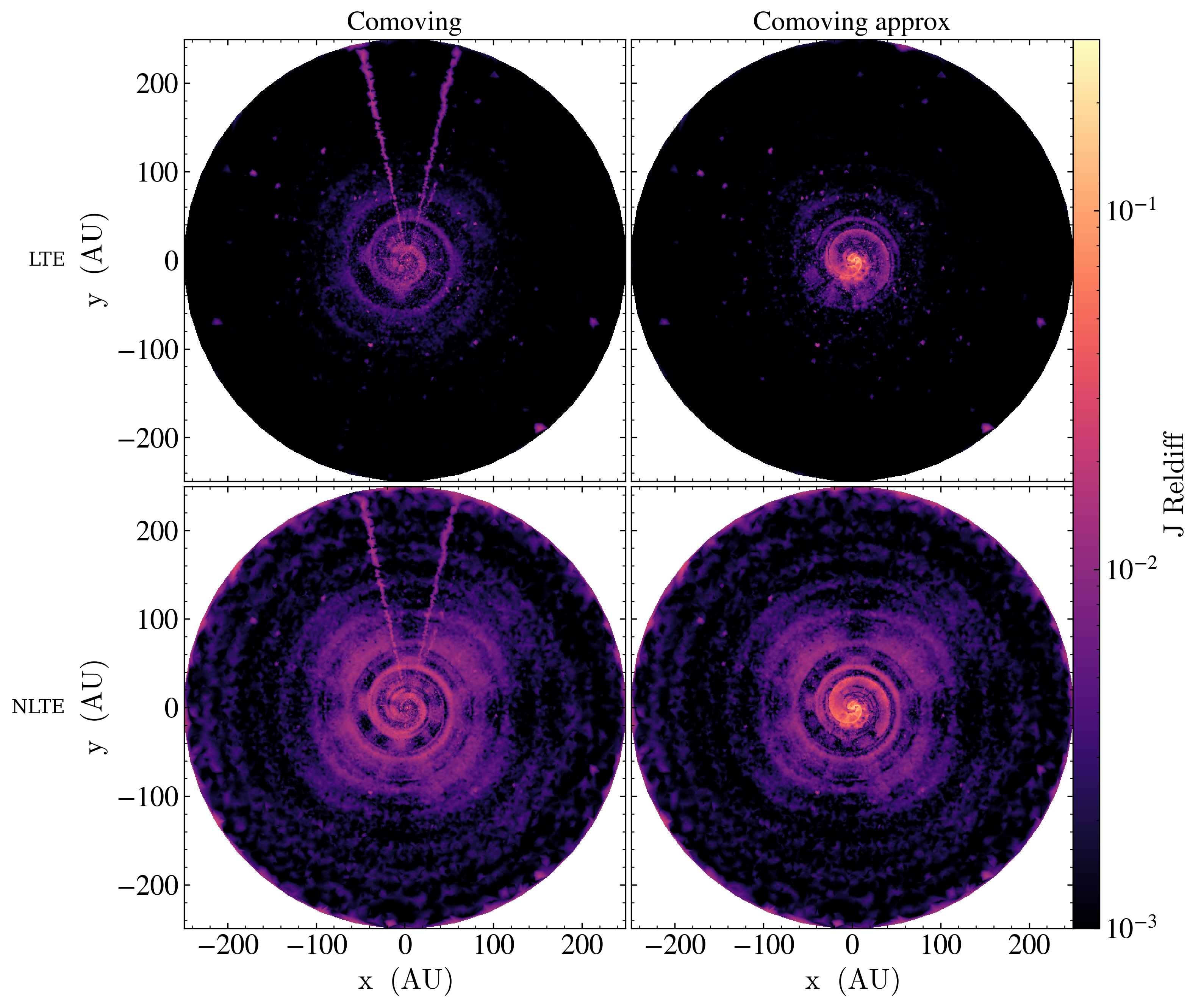}
\caption{Slice plots of the relative differences between the computed mean line intensity $J_{ul}$ for the CO J=2-1 v=0 line using the Feautrier solver and both comoving solvers. The relative differences in mean line intensity have been calculated both after a single iteration starting from LTE (LTE) and after convergence (NLTE).
}\label{figure: Relative diff J NLTE image}
\end{figure}

\begin{figure}
	\centering
	\includegraphics[width=\linewidth]{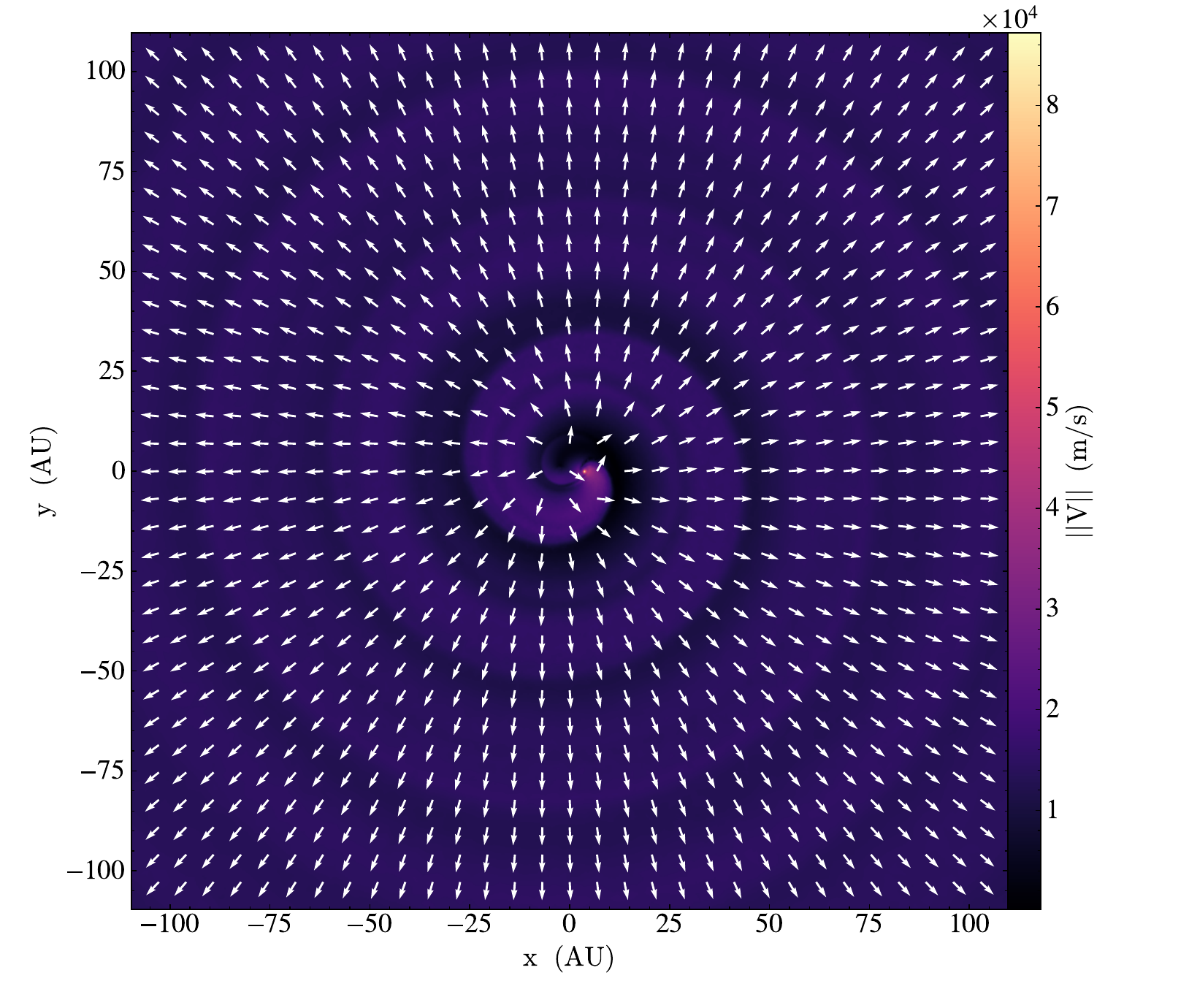}
	\caption{Zoomed in slice plot of the velocity field of the \textsc{Phantom} hydrodynamics model. The arrows denote the direction of the velocity field.}\label{figure: velocity field}
\end{figure}

\begin{figure}
	\includegraphics[width=\linewidth]{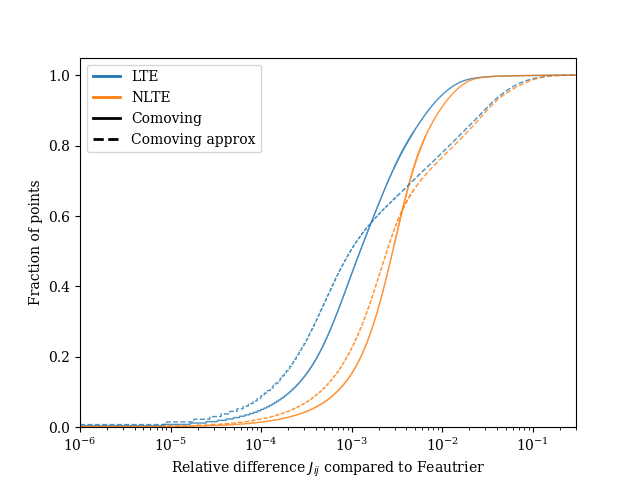}
	\caption{Cumulative difference between the computed mean line intensity $J_{ul}$ for the CO J=2-1 v=0 line using the Feautrier solver and the comoving solvers with an adaptive angular discretization.  Both the differences in case of a single iteration after setting the level populations $n_i$ to LTE (in blue) and the difference after letting the level populations converge (in orange). The figure shows the results for the correct boundary conditions described in \ref{Appendix: boundary condition implementation} as solid lines and the results when using the approximate boundary conditions as dashed lines.}\label{figure: comparison differences comoving implementations}
\end{figure}

In order to be able to compare the comoving solver of this paper and the reference Feautrier method without the presence of numerical artifacts, we implement an adaptive angular discretization, as explained in Section \ref{section: adaptive angular discretization}.
As seen in Fig. \ref{figure: Relative diff J NLTE image}, both versions of the comoving solver perform well in the outside regions of the model. But, the comoving solver version with more correct treatment of the boundary, is more accurate in the center of the model, where the velocity field is non-monotonic (see Fig. \ref{figure: velocity field}). We remark that we have not been able to fully eliminate the numerical artifacts seen in Section \ref{section: static angular discretization application}. We think this to be caused due to the rays used in the comoving solver not passing exactly through each model point, which slightly changes the viewing angle towards the dense regions. The numerical artifacts are not occurring in the approximate version of the comoving solver because the intensity information from the dense center gets lost due to Doppler shifts.
In Fig. \ref{figure: comparison differences comoving implementations}, we see that the relative differences of the mean line intensity $J_{ul}$ at NLTE, compared to the value computed using the Feautrier solver, lie below $1\%$ for $94\%$ of the points in the model in case of the comoving solver, and below $1\%$ for $78\%$ of the points in the model in case of the comoving solver with approximate boundary conditions. 
We find that the adaptive angular discretization has a significant impact on the computational performance, due to decreased reuse of the computed intensities $I(\hat{\boldsymbol{n}}, \boldsymbol{x}, \nu)$ on rays, as different positions now require a different set of directions. As shown in Table \ref{table: timings NLTE}, we need relatively more rays to compute the entire intensity field, compared to the uniform angular discretization case (about $10\%$ of the rays used by the Feautrier solver compared to $5\%$ when using a uniform angular discretization). This results in an insignificant speedup for the comoving solver with correct boundary conditions, and a significantly lower speedup for the comoving solver with approximate boundary conditions, compared to Section \ref{section: static angular discretization application}.

In this model, which has a spherically symmetric outflow perturbed by a companion, it appears that the more correct boundary conditions are unnecessary for the outer regions of the model. In these areas, the velocity field can be approximated by a monotonic profile for most directions of the entire solid angle. The only region where the more accurate boundary conditions are required, is the very center of the model. In all other regions of the model, disregarding the non-monotonic behavior of the velocity field may result in insignificant errors, potentially allowing us to save computation time (compare both versions of the comoving solver in Table \ref{table: timings NLTE}).\\

\begin{table}
	\centering
	\caption{Time required to compute NLTE level populations (5 iterations) on the \textsc{Phantom} model, containing $885501$ points, using an adaptive angular discretization. To illustrate the maximum possible obtainable speedup, we also add the number of rays used to compute the intensity field per solver. Do note that this theoretical speedup assumes that all computation methods require the same computation time per ray.  This table compares the Feautrier solver, the comoving solver with correct boundary conditions, and the comoving solver with approximate boundary conditions.}
	\label{table: timings NLTE}
	\begin{tabular}{l|r|r}
		Method&Time[s]&Rays[.]\\
		\hline
		Feautrier& $1.9\cdot 10^4$&$3.2\cdot 10^7$\\
		Comoving & $1.8 \cdot 10^4$&$3.1\cdot 10^6$\\
		Comoving approx& $7.2 \cdot 10^3$&$3.1\cdot 10^6$\\
	\end{tabular}
\end{table}

\section{Conclusion}

In this paper, we have derived a computationally efficient, numerically stable method for calculating the radiation field in line radiative transfer by enabling reuse of previously computed intensities. We derive the computation method in Section \ref{section: comoving solver}, where we have discretized a version of the 1D radiative transfer equation which allows for a change in frequency, similar to \citep{baron_co-moving_2004,hauschildt_improved_2004}. We prove that the numerical stability of the method mostly depends on the ratio of the Doppler shift versus the spacing of the frequency discretization (see Section \ref{section: stability analysis}). To improve the numerical stability of the discretization, we introduce a technique we call frequency matching, detailed in Section \ref{section: Frequency matching}. In this technique, we ensure that the frequency paths $\nu_i(x)$ used for discretizing the equations cover the shortest possible distance in frequency space. Through a simple 1D plane-parallel example (see Sections \ref{section: instabilities}, \ref{section: frequency matching explanation}), we show that this modification can significantly reduce the oscillatory behavior of the resulting solver. This frequency matching procedure introduces additional boundary conditions in our numerical discretization, which we treat in Section \ref{section: boundary conditions}. The importance of using an appropriate implementation for the boundary condition is demonstrated in Section \ref{section: comoving benchmark}.
In the first implementation, the previously computed intensities are stored and interpolated to obtain an accurate estimate for intensities at later points, while the second implementation always assumes a black-body intensity.
Afterwards, we apply this computation method to an actual 3D hydrodynamics model in Section \ref{section: application} in order to assess both its accuracy and speedup compared to the Feautrier solver \citep{feautrier_sur_1964}, using two different treatments for the boundary condition. 
In our application, we noticed numerical artifacts appearing in the results for all solvers, due to the uniform angular discretization we used to compute the mean intensity. Therefore, we devised a method to adaptively sample the angular discretization of the solid angle to eliminate these numerical effects (see Section \ref{section: adaptive angular discretization}). It should be noted, however, that this new angular discretization results in reducing the computational efficiency for the comoving solvers by a factor $2$, due to the more limited reuse of the computed intensity. The comoving solver with more accurate boundary conditions is more precise than the other implementation, especially in the inner, more turbulent regions of the model, but is also significantly slower than the more approximate implementation. 
The former achieves speedup of a factor $2$ compared to the Feautrier solver when using a uniform angular discretization, but this speedup becomes negligible when using an adaptive angular discretization. Similarly, the latter implementation obtains a speedup of a factor $5$, which drops to a factor $2.5$ when an adaptive angular discretization is used.\\

\section*{Acknowledgments}
T. Ceulemans has been supported by the Research Foundation - Flanders (FWO) grant 1166722N and 1166724N. F. De Ceuster acknowlegdes support from the Research Foundation - Flanders (FWO), grants 1253223N and I002123N. Leen Decin acknowledges support from the KU Leuven C1 MAESTRO grant C16/23/009 and from the KU Leuven methusalem SOUL grant METH/24/012.
The authors express their thanks to J. Malfait for providing the \textsc{Phantom} model used in this paper. The authors thank the anonymous referee for their comments, which significantly helped in improving the clarity of this manuscript.

\section*{Data Availability}

\textsc{Magritte} is an open-source software library available at \url{https://github.com/Magritte-code/Magritte}. The code used to generate the results and figures of this paper can be found in the fork at \url{https://github.com/Magritte-code/Magritte_Paper_Comoving}.

\appendix

\section{Boundary condition implementation}\label{Appendix: boundary condition implementation}
For the comoving solver, we require extra information on outermost frequencies of the frequency discretization. These are boundary conditions. If only line radiative transfer is considered (and no continuum sources, such as dust), the intensities at frequencies far away from the line centers, can be considered constant. One possibilty is to use the initial intensities, computed at the start of the ray, as boundary conditions. To simplify the explanation later on, we call these initial boundary conditions.
At other places, we might require an intensity at a frequency we have encountered before, but have doppler shifted away from (see Fig. \ref{figure: boundary conditions}). This can happen because of a non-monotonic velocity field. In the next part, these are called returning boundary conditions.

As implementing both frequency matching (as defined in Section \ref{section: Frequency matching}) and appropriate boundary conditions together in a numerical code is complicated\footnote{The authors are not aware of any other literature describing a numerical method with similar implementation challenges. Therefore we consider this algorithm to be a first implementation attempt, which can be further optimized.}, we describe in this appendix how we implemented this in \textsc{Magritte}\footnote{Available at \url{https://github.com/Magritte-code/Magritte}} \cite{ceulemans_magritte_2024, de_ceuster_3d_2022, deceuster_magritte_2020, deceuster_magritte_2019}.

\subsection{The philosophy behind the implementation}
When computing the intensity using the comoving solver, one does not know in advance which intensities (at a specific frequency and position) will be required by a returning boundary condition. To prevent the excessive use of if-clauses and to simplify the code a bit, we will first map all data needed (source functions $S_{j,i}$, optical depth increments $\Delta\tau$, frequency derivative coefficients $c_{ijk}$, and position index\footnote{The position index is required for the practical implementation of boundary conditions. If we happen to Doppler shift away from and back to a specific frequency during the computations on a single ray, we need to interpolate based on intensities computed from the previous position.} to start from) to do the computations on a single ray, before computing any intensity. It turns out that the general discretization of the co-moving frame formal solution (see Eq. \ref{eq: moving solver discretization}) is flexible enough for us to implement boundary conditions using the same computations. We will therefore exploit the formulation to make computation consistent.\\

For an initial boundary condition, one can put the frequency derivative coefficients $c_{ijk}$ to $0$, the optical depth $\Delta\tau_i$ very large (e.g. $\Delta\tau_i=50$) and the source functions $S_{j,i}, S_{j+1,i}$ equal to the boundary intensity $I_{\text{bdy}}$. In this way, the computed intensity will be approximately equal to the boundary intensity $I_{\text{bdy}}$:
\begin{align}
	I_{j+1,i} = I_{j,i}e^{-\Delta\tau_i} + I_{\text{bdy}}\left(1-e^{-\Delta\tau_i}\right) \simeq I_{\text{bdy}}.
\end{align}

In case of a boundary condition for which we already have computed intensities at frequencies around the requested frequency, we can interpolate these previous results by noticing that the required frequency lies between two frequencies $\nu_i<\nu_{i+1}$ from which we shifted away on previous points on the ray. Then we linearly interpolate those previous intensities in order to obtain a boundary intensity. This can be done by pointing the computation data towards using the intensities at those frequencies at the corresponding previous points, and setting the frequency derivative coefficients $c_{ijk}$ for the explicit part of the computations to $[-2/(\nu_{i+1}(x_j)-\nu_i(x_j)), 2/(\nu_{i+1}(x_j)-\nu_i(x_j)),0]$, for the implicit part to $0$ and setting the optical depth $\Delta\tau_i$ to almost zero\footnote{Using $\Delta\tau_i=0$ is not recommended, due to dividing by zero in both the source and shift term of Eq. \ref{eq: moving solver discretization} Note that in the limit $\Delta\tau_i\to 0$, both terms converge to a limit value, thus it is possible to evaluate these terms using Taylor expansions when $\Delta\tau_i$ close to $0$.}, e.g. $\Delta\tau_i = 10^{-10}$. The source functions $S_{j,i}, S_{j+1,i}$ will be put to $0$. The resulting equation will interpolate the intensity:
\begin{align}
I_{j+1,i} &= I_{j,i}e^{-\Delta\tau_i}
+\frac{\Delta \nu_{\text{obs},i}}{\Delta\tau_i}\frac{2I_{j, i+1}-2I_{j, i}}{\nu_{i+1}(x_j)-\nu_i(x_j)}\left(\frac{1-e^{-\Delta\tau_i}-\Delta\tau_i e^{-\Delta\tau_i}}{\Delta\tau_i}\right)\nonumber\\
&\simeq I_{j,i} + \frac{\Delta\nu_{\text{obs},i}}{\nu_{i+1}(x_j)-\nu_i(x_j)}\left(I_{j, i+1}-I_{j, i}\right)
\end{align}

\subsection{Determining the frequency boundary points}

Now we still need to decide when to use boundary conditions, and to which type they belong. 
Assume that we want to compute intensities starting from the start of a ray, until we reach the end. For determining the boundary conditions, we will start from the last point of the ray instead. First, we note down the frequency range which is spanned by each frequency quadrature of all lines, excluding the boundary points required for the discretization in frequency space and store these boundary frequencies in a map\footnote{The map type should allow for duplicate keys, as it can happen that the required frequencies at different spatial positions may overlap (e.g. in case of no velocity difference between successive points). In C++, we use the std::multimap for this.}, using frequency as key and spatial position index as value. Second, we compare the currently stored map of boundary points against the currently spanned frequency ranges, if any boundary frequencies are contained, then (at intensity calculation time) these are returning boundary conditions and should be interpolated using the encompassing frequencies at the current spatial position. If so, they can be removed from the map of still to match boundary frequencies. This procedure is then repeated at all previous spatial positions of the ray, from end to start. Afterwards, if any boundary points remain in the map, they are then treated as initial boundary conditions.

\section{Discretization of a second order accurate frequency derivative}\label{appendix: coefficients second order freq derivative}
For a second order accurate frequency derivative in $\nu$, we need to derive the quadrature coefficients. These are found by solving the following system of equations
\begin{equation}
\begin{cases}
\sum_{k\in\{0,1,2\}} c_{ijk}=0\\
\sum_{k\in\{0,1,2\}} c_{ijk}(\nu_{i+k}(x_j)-\nu_i(x_j))=1\\
\sum_{k\in\{0,1,2\}} c_{ijk}(\nu_{i+k}(x_j)-\nu_i(x_j))^2=0
\end{cases}
\end{equation}
Solving the third equation, we get $$c_{ij1}=-c_{ij2}\frac{(\nu_{i+2}(x_j)-\nu_i(x_j))^2}{(\nu_{i+1}(x_j)-\nu_i(x_j))^2}.$$ 
This, we fill in into the second equation $$c_{ij2}\left((\nu_{i+2}(x_j)-\nu_i(x_j))-\frac{(\nu_{i+2}(x_j)-\nu_i(x_j))^2}{\nu_{i+1}(x_j)-\nu_i(x_j)}\right)=1$$ and we find:
$$c_{ij2}=\frac{(\nu_{i+1}(x_j)-\nu_i(x_j))}{(\nu_{i+2}(x_j)-\nu_i(x_j))(\nu_{i+1}(x_j)-\nu_i(x_j))-(\nu_{i+2}(x_j)-\nu_i(x_j))^2}$$
Thus $c_{ij1}$ is given by
$$c_{ij1}=\frac{(\nu_{i+2}(x_j)-\nu_i(x_j))}{(\nu_{i+2}(x_j)-\nu_i(x_j))(\nu_{i+1}(x_j)-\nu_i(x_j))-(\nu_{i+1}(x_j)-\nu_i(x_j))^2}$$
The explicit form\footnote{The explicit form is only useful for derivations; use $c_{ij0}=-c_{ij1}-c_{ij2}$ for any practical implementation.} of $c_{ij0}$ is given by
$$c_{ij0}=\frac{1}{\nu_{i+2}(x_j)-\nu_{i+1}(x_j)}\left(\frac{\nu_{i+1}(x_j)-\nu_i(x_j)}{\nu_{i+2}(x_j)-\nu_i(x_j)}-\frac{\nu_{i+2}(x_j)-\nu_i(x_j)}{\nu_{i+1}(x_j)-\nu_i(x_j)}\right)$$
and is always negative if $\nu_i(x_j)<\nu_{i+1}(x_j)<\nu_{i+2}(x_j)$.



\bibliographystyle{elsarticle-harv} 
\bibliography{Library_bibtex_20250218}





\end{document}